\documentclass[twocolumn, prb, amsfonts, amsmath, amssymb, superscriptaddress, floatfix, aps, a4paper, 10pt]{revtex4-2}
\usepackage[T1]{fontenc}
\usepackage{lmodern}
\usepackage{graphicx}
\usepackage[dvipsnames]{xcolor}
\usepackage[colorlinks,linkcolor=Blue,citecolor=Blue]{hyperref}
\usepackage{braket}
\usepackage{mathtools}
\usepackage{microtype}

\begin{document}

\title{Algebraic Exact Solution for Driven Landau Levels in Two-dimensional Electron Gases}

\author{Li-kun Shi}
\affiliation{Institut f{\"u}r Theoretische Physik, Universit{\"a}t Leipzig, Br{\"u}derstra{\ss}e 16, 04103, Leipzig, Germany}
\affiliation{Center for Quantum Matter, School of Physics, Zhejiang University, Hangzhou 310058, China}

\date{\today}

\begin{abstract}
Controlling quantum systems with time-dependent fields opens avenues for engineering novel states of matter and exploring non-equilibrium phenomena. Landau levels in two-dimensional electron gases (2DEGs), with their discrete energy spectrum and characteristic cyclotron dynamics, provide an important platform for realizing and studying such driven quantum systems. While exact solutions for driven Landau levels exist, they have been limited to specific gauges or representations. In this work, we present an algebraic, gauge- and representation-independent exact solution for driven Landau levels in 2DEGs subject to arbitrary time-dependent electromagnetic fields. Our approach, based on a time-dependent unitary transformation via the displacement operator, provides clear physical insights into the driven quantum dynamics. We apply this method to derive the exact Floquet states and quasienergies for periodically driven Landau levels, and we extend our analysis to the resonant driving regime, where the Floquet picture breaks down and the electron wavefunction exhibits unbounded spatial spreading. Furthermore, we calculate the instantaneous energy absorption rate, revealing distinct absorption behaviors between coherent states and Fock or thermal states, stemming from quantum interference effects.
\end{abstract}

\maketitle
\section{Introduction}

The control of quantum systems through time-dependent fields represents an active area of research in condensed matter physics, offering possibilities for exploring non-equilibrium phenomena and quantum dynamics. Floquet engineering, where periodic driving modifies effective Hamiltonians, has emerged as one approach to studying these systems~\cite{bukov2015universal, eckardt2017colloquium, oka2019floquet}.
Two-dimensional electron gases (2DEGs) in magnetic fields provide an important platform for studying driven quantum systems. The magnetic field quantizes the electron motion into Landau levels with a characteristic cyclotron frequency $\omega_c$. These systems are experimentally accessible in semiconductor heterostructures and graphene, where the electron dynamics can be probed through transport and optical measurements~\cite{von1986quantized, dmitriev2012nonequilibrium, McIver2020}.
While many studies focus on high frequencies, the behavior of these systems under strong driving fields or at low frequencies represents a less explored regime.

Current theoretical approaches face certain limitations in analyzing driven Landau levels. Time-dependent perturbation theory and the Floquet-Magnus expansion~\cite{grifoni1998driven} are most applicable when the driving is weak and at high frequencies. Foundational work has established exact solution methods for this class of problems, such as the theory of time-dependent invariants~\cite{lewis1969exact}. However, many existing explicit solutions have been derived in specific gauges (e.g., the Landau gauge) or representations (e.g., real-space wavefunctions)~\cite{Husimi1953, popov1969parametric, popov1970parametric, dittrich1998quantum, inoshita2000light, dini2016magnetic, herath2022generalized}. The gauge-dependent nature of these solutions presents inherent limitations, particularly for problems where specific gauge choices obscure the underlying physics.

Here, we attempt to contribute to this field by presenting an algebraic approach to solving driven Landau levels that maintains gauge independence. Our method utilizes a time-dependent unitary transformation based on the displacement operator, drawing inspiration from techniques in quantum optics~\cite{scully1997quantum}. In Section~\ref{sec:exact-2DEG}, we develop this framework and derive the solution. Section~\ref{sec:Application} explores potential applications, including an analysis of on-resonant conditions. Section~\ref{sec:discussion} summarizes our findings and discusses potential extensions.

\section{Exact Solution for Driven Landau Levels in 2DEG}
\label{sec:exact-2DEG}

We begin by considering a two-dimensional electron gas (2DEG) in the $xy$-plane, subject to both a perpendicular magnetic field ${\bf B} = B\hat{z}$ and a time-dependent vector potential ${\bf A}(t)$. The Hamiltonian describing this system takes the form:
\begin{align} 
\begin{aligned}
\hat{H}(t) = \frac{1}{2m} \left( \hat{\bf p} - e{\bf A}_0 (\hat{\bf r}) - e{\bf A}(t) \right)^2,
\label{eq:hamiltonian}
\end{aligned} 
\end{align}
where $\hat{\bf p} = (\hat{p}_x, \hat{p}_y)$ is the momentum, $m$ is the effective mass, and ${\bf A}_0 (\hat{\bf r})$ is the static vector potential generating the magnetic field $B\hat{z} = \nabla \times {\bf A}_0 (\hat{\bf r})$.

We first introduce the canonical momentum operators,
\begin{align} 
\begin{aligned}
\hat{\pi}_x = \hat{p}_x - eA_{0x} (\hat{\bf r}),
\quad
\hat{\pi}_y = \hat{p}_y - eA_{0y} (\hat{\bf r}) .
\label{eq:canonical}
\end{aligned} 
\end{align}
They satisfy the commutation relation independent of gauge choices:
\begin{align} 
\begin{aligned}
[\hat{\pi}_x, \hat{\pi}_y] = -i\hbar eB .
\end{aligned} 
\end{align}
With the magnetic length $l_B = \sqrt{\hbar/eB}$ and cyclotron frequency $\omega_c = eB/m$, we define the standard ladder operators:
\begin{align} 
\begin{aligned}
\hat{a} = \frac{l_B}{\sqrt{2}\hbar}(\hat{\pi}_x - i\hat{\pi}_y),
\quad
\hat{a}^\dagger = \frac{l_B}{\sqrt{2}\hbar}(\hat{\pi}_x + i\hat{\pi}_y),
\label{eq:ladder_ops}
\end{aligned} 
\end{align}
which satisfy $[\hat{a}, \hat{a}^\dagger] = 1$.

We consider a time-dependent vector potential of the form:
\begin{align} 
\begin{aligned}
{\bf A}(t) = [A_x (t), A_y (t)],
\quad
A_{x,y} (t) \in \mathbb{R}
\label{eq:vector_potential}
\end{aligned} 
\end{align}
with ${\bf E}(t) = - \partial_t {\bf A}(t)$ being the physical electric field.
Expressing the Hamiltonian in terms of these ladder operators yields:
\begin{align} 
\begin{aligned}
\hat{H}(t) &= \hbar\omega_c\left(\hat{a}^\dagger \hat{a} + \frac{1}{2}\right) - \hat{a} z_t^* - \hat{a}^\dagger z_t + 
\frac{e^2{\bf A}^2 (t)}{2m},
\label{eq:hamiltonian_ladder}
\end{aligned} 
\end{align}
where we have introduced the complex driving term which has a unit of energy:
\begin{align} 
\begin{aligned}
z_t & = \sqrt{\frac{\hbar \omega_c}{2 m}} [ eA_x(t) - ieA_y(t) ] .
\label{eq:driving_term}
\end{aligned} 
\end{align}
This form of the Hamiltonian clearly separates the static Landau level part (first term) from the driving terms (second and third terms) and the time-dependent scalar potential (last term).

To solve the time-dependent Schrödinger equation
\begin{align} 
\begin{aligned}
i\hbar
\frac{\partial}{\partial t}  \ket{\psi(t)} = \hat{H}(t)\ket{\psi(t)},
\label{eq:schrodinger}
\end{aligned} 
\end{align}
we employ a time-dependent unitary transformation.  Our strategy is to move to a frame that ``follows'' the classical motion induced by the driving field.  This is analogous to the classical solution of a driven harmonic oscillator, where one shifts the coordinate system to the time-dependent equilibrium point.  In the quantum case, the appropriate transformation is generated by the displacement operator~\cite{scully1997quantum}:
\begin{align} 
\begin{aligned}
\hat{D}(\alpha_t) = \exp\left[\alpha_t \hat{a}^\dagger - \alpha_t^* \hat{a}\right], \quad
\hat{D}^\dagger (\alpha_t) \hat{D} (\alpha_t) = 1,
\label{eq:displacement}
\end{aligned} 
\end{align}
where $\alpha_t$ is a time-dependent complex function to be determined.  Physically, $\alpha_t$ represents the complex displacement of the electron's guiding center in phase space, mirroring the classical trajectory driven by the external field, given by:
\begin{align}
\begin{aligned}
\hat{D}^\dagger(\alpha_t)\hat{a}\hat{D}(\alpha_t) &= \hat{a} + \alpha_t, \\
\hat{D}^\dagger(\alpha_t)\hat{a}^\dagger \hat{D}(\alpha_t) &= \hat{a}^\dagger + \alpha_t^*.
\label{eq:transformed_ops}
\end{aligned}
\end{align}
The transformed state vector
\begin{align} 
\begin{aligned}
\ket{\psi'(t)} = \hat{D}^\dagger(\alpha_t)\ket{\psi(t)},
\label{eq:transformed_state}
\end{aligned} 
\end{align}
evolves according to the transformed Hamiltonian $\hat{H}'(t) $:
\begin{align} 
\begin{aligned}
& i\hbar
\frac{\partial}{\partial t} \ket{\psi'(t)}
= \hat{H}'(t)\ket{\psi'(t)}, \\
& \hat{H}'(t) = \hat{D}^\dagger(\alpha_t)\hat{H}(t) \hat{D}(\alpha_t) - i\hbar \hat{D}^\dagger(\alpha_t)
\frac{\partial}{\partial t} \hat{D}(\alpha_t) .
\label{eq:transformed_evolution}
\end{aligned} 
\end{align}
Using the properties of the displacement operator (see Appendix~\ref{app:displacement_derivative}), the transformed Hamiltonian becomes:
\begin{align}
& \hat{H}'(t) = \hbar\omega_c\left(\hat{a}^\dagger \hat{a} + \frac{1}{2}\right) + \gamma_t \hat{a}^\dagger + \gamma_t^* \hat{a}
+ \Delta_t,
\label{eq:Hprime-2DEG}
\end{align}
where
\begin{align}
\gamma_t = \hbar\omega_c \alpha_t - z_t - i\hbar\dot{\alpha}_t ,
\end{align}
and $\Delta_t = \hbar\omega_c |\alpha_t|^2 - 2 {\rm Re}[\alpha_t z_t^*] - \hbar{\rm Im}[\alpha_t\dot{\alpha}_t^*] + e^2{\bf A}^2 (t)/2m$ is a time-dependent scalar function.

To eliminate the time-dependent terms linear in $\hat{a}$ and $\hat{a}^\dagger$, we need to solve for $\alpha_t^{(0)}$ that satisfies:
\begin{align} 
\begin{aligned}
\gamma_t^{(0)} = \hbar \omega_c \alpha_t^{(0)} - z_t - i\hbar\dot{\alpha}_t^{(0)} = 0.
\label{eq:alpha_condition}
\end{aligned} 
\end{align}
This condition ensures that the transformed Hamiltonian describes a simple, undriven harmonic oscillator (plus a time-dependent scalar potential).  Intuitively, we are choosing the displacement $\alpha_t^{(0)}$ to precisely cancel out the driving force in the transformed frame.

With the solution for $\alpha_t^{(0)}$, the transformed Hamiltonian reduces to:
\begin{align} 
\begin{aligned}
\hat{H}'(t) = \hbar\omega_c\left(\hat{a}^\dagger \hat{a} + \frac{1}{2}\right) + \Delta_t^{(0)},
\label{eq:final_transformed_H}
\end{aligned} 
\end{align}
where 
\begin{align} 
\begin{aligned}
\Delta_t^{(0)} & = \hbar\omega_c |\alpha_t^{(0)}|^2 - 2 {\rm Re}\big(\alpha_t^{(0)} z_t^* \big) 
\\
& \quad - \hbar{\rm Im} \big[ \alpha_t^{(0)} \big( \dot{\alpha}_t^{(0)} \big)^* \big] + 
\frac{e^2{\bf A}^2 (t)}{2m} .
\end{aligned} 
\end{align}
The solution to the transformed Schrödinger equation Eq.~\eqref{eq:transformed_evolution} is then:
\begin{align}
\begin{aligned}
\ket{N'(t)} =
e^{-\frac{i}{\hbar}\int_{t_0}^t \Delta_{t'}^{(0)}dt'}
e^{-i \omega_c (\hat{a}^\dagger \hat{a} +1/2)  t}
\ket{N},
\label{eq:transformed_solution}
\end{aligned}
\end{align}
where $\ket{N}$ are the static Landau level states. The exact solution in the original frame is obtained by applying the inverse transformation:
\begin{align}
\begin{aligned}
& \ket{N(t)} = \hat{D}(\alpha_t^{(0)})\ket{N'(t)} \\
&= e^{-\frac{i}{\hbar}\int_{t_0}^t \Delta_{t'}^{(0)}dt'} \hat{D}(\alpha_t^{(0)})
e^{-i \omega_c (\hat{a}^\dagger \hat{a} +1/2)  t}
\ket{N}.
\label{eq:exact_solution}
\end{aligned}
\end{align}

This solution is the central result of our paper. It is expressed entirely in terms of the ladder operators $\hat{a}$ and $\hat{a}^{\dagger}$, and the displacement parameter $\alpha_t^{(0)}$, which is determined by the driving field $z_t$. This algebraic form makes the solution independent of specific choice of gauge or representation. The dynamics are encoded in the time-dependent displacement of the Landau level states, reflecting the classical motion induced by the driving.

\section{Application of Exact Solution}
\label{sec:Application}

We now demonstrate the practical applications of our solution for driven Landau levels. To validate our approach, we first examine the simplest case: periodic driving at frequencies away from the cyclotron frequency. This allows direct comparison with established results.

We then analyze the case where the driving frequency matches the cyclotron frequency. In this resonant regime, the standard Floquet description breaks down. Here, our method reveals new insights about the system's non-equilibrium behavior.

While our solution is gauge-independent, practical calculations sometimes benefit from choosing a specific gauge. For example, when computing position expectation values, selecting an appropriate gauge can simplify the calculation without compromising the generality of our results.

\subsection{Non-resonant Floquet Landau Levels in 2DEG}
\label{subsec:non-resonant-Floquet-2DEG}

We first consider a time-dependent driving vector potential of the form:
\begin{align} 
\begin{aligned}
{\bf A}(t) = [A_x \sin(\Omega t), A_y \sin(\Omega t + \phi_y)],
\quad
A_{x,y} \in \mathbb{R}
\label{eq:vector_potential_floquet}
\end{aligned} 
\end{align}
Here, $\Omega \neq \omega_c$ represents the driving frequency (with period $T = 2\pi/\Omega$), and $\phi_y$ introduces a phase delay between the $x$ and $y$ components, allowing for different polarization patterns of the driving field.

This periodic driving transforms our earlier complex driving terms from Eq.~\eqref{eq:driving_term} into frequency components at $\pm\Omega$:
\begin{align} 
\begin{aligned}
& z_t = z_+ e^{-i\Omega t} + z_- e^{+i\Omega t},
\\
& z_+ = \sqrt{\frac{\hbar \omega_c}{2 m}}
\left(-\frac{e A_x}{2i} + \frac{e A_y}{2}e^{-i\phi_y} \right),\\
& z_- = \sqrt{\frac{\hbar \omega_c}{2 m}} \left(+\frac{e A_x}{2i} - \frac{e A_y}{2}e^{+i\phi_y} \right) ,
\label{eq:driving_term_floquet}
\end{aligned} 
\end{align}

To find the exact solution for this periodic driving, we need to solve Eq.~\eqref{eq:alpha_condition} that determines $\alpha_t^{(0)}$. Given the periodic nature of our driving, we make an ansatz that $\alpha_t^{(0)}$ should have the same frequency components as the driving:
\begin{align} 
\begin{aligned}
\alpha_t^{(0)} = \alpha_+^{(0)}e^{-i\Omega t} + \alpha_-^{(0)}e^{+i\Omega t}
\end{aligned} 
\end{align}
Substituting this ansatz into Eq.~\eqref{eq:alpha_condition} and matching frequency components, we obtain:
\begin{align} 
\begin{aligned}
\alpha_+^{(0)} = \frac{z_+}{\hbar(\omega_c - \Omega)}, 
\quad
\alpha_-^{(0)} = \frac{z_-}{\hbar(\omega_c + \Omega)}.
\label{eq:alpha_pm}
\end{aligned} 
\end{align}
With these solutions, we can evaluate the time-dependent scalar term $\Delta_t^{(0)}$, which separates naturally into a static part $\Delta_0$ and an oscillating part:
\begin{align} 
\begin{aligned}
\Delta_0^{(0)} = -\frac{|z_+|^2}{\hbar(\omega_c-\Omega)} -\frac{|z_-|^2}{\hbar(\omega_c+\Omega)} + \frac{e^2(A_x^2 + A_y^2)}{4m}.
\label{eq:Delta_0}
\end{aligned} 
\end{align}
The oscillating part takes the form:
\begin{align}
& \Delta_{\text{osc}}^{(0)} (t) = \beta e^{-2i\Omega t} + \beta^* e^{+ 2i\Omega t},
\\
& \beta = \hbar \omega_c \frac{z_+ z_-^*}{\hbar^2(\omega_c - \Omega)(\omega_c + \Omega)}  +  \frac{z_+ z_-^*}{\hbar \omega_c} .
\end{align}
These expressions allow us to write our solution $\ket{N(t)}$ in Eq.~\eqref{eq:exact_solution} in the standard Floquet form:
\begin{align} 
\begin{aligned}
& \ket{N(t)} = e^{-i \varepsilon_N t / \hbar} \ket{\phi_N(t)}, \\
& \ket{\phi_N(t)} = \ket{\phi_N(t+T)},
\label{eq:floquet_form}
\end{aligned} 
\end{align}
where the quasienergies are given by~\cite{dini2016magnetic, herath2022generalized}:
\begin{align} 
\begin{aligned}
\varepsilon_N = \hbar \omega_c \left(N + \frac{1}{2}\right) + \Delta_0^{(0)}.
\label{eq:quasienergy}
\end{aligned} 
\end{align}
We note that when the driving frequency approaches the cyclotron frequency ($\Omega  \to \pm \omega_c$), the complex displacement $\alpha_t^{(0)}$ and the static shift $\Delta_0^{(0)}$ diverges [see Eqs.~\eqref{eq:alpha_pm} and \eqref{eq:Delta_0}]. This divergence signals a breakdown of our solution $\alpha_t^{(0)}$ at resonance.

However, we will show in the next section, the more generic exact solution from Eq.~\eqref{eq:alpha_condition} still works at resonant driving frequency.

\subsection{Resonant Driving: Beyond Floquet Dynamics}
\label{subsec:resonant-floquet-2deg}

We now turn to the case of resonant driving, where the driving frequency $\Omega$ equals the cyclotron frequency $\omega_c$. This regime is of particular interest because it is expected to lead to significant energy transfer and a breakdown of the Floquet picture. While Popov~\cite{popov1970parametric} briefly suggested, without detailed mathematical analysis, that resonant driving induces a transition from a discrete to a continuous energy spectrum, the full nature of this transition remains to be thoroughly investigated. Here, we aim to provide a detailed analysis of this resonant regime using our exact algebraic approach.

We focus specifically on the case where $\Omega = \omega_c$ (the $\Omega = - \omega_c$ case can be similarly obtained by exchanging $z_{-}$and $z_{+}$ and replacing $ \omega_c$ with $-\omega_c$).  In this resonant condition, the time-dependent driving term takes the form [see Eq.~\eqref{eq:driving_term_floquet}]:
\begin{align}
z_t = z_+ e^{-i\omega_c t} + z_- e^{+i\omega_c t},
\label{eq:z_t_resonant}
\end{align}
leading to the following equation for $\alpha_t^{(0)}$:
\begin{align}
i\hbar\dot{\alpha}_t^{(0)} - \hbar \omega_c \alpha_t^{(0)} = - (z_+ e^{-i\omega_c t} + z_- e^{+i\omega_c t}).
\label{eq:alpha_resonance}
\end{align}
Assuming $\alpha_{t=0}^{(0)} = 0$ (i.e., the system starts at rest), the solution to this linear first-order differential equation is:
\begin{align}
\alpha_t^{(0)} = + i t \frac{z_+}{\hbar}e^{-i\omega_c t} 
+ i \frac{z_-}{\hbar \omega_c} \sin{\omega_c t} .
\label{eq:alpha_resonance_solution}
\end{align}
The solution for $\alpha_t^{(0)}$ contains a term that grows linearly with time, indicating that the displacement of the Landau level states in phase space increases without bound as time progresses.  Classically, a driven harmonic oscillator at resonance exhibits unbounded amplitude growth.  In our quantum system, this corresponds to the electron's wavefunction becoming increasingly delocalized.

To illustrate this delocalization, we calculate the expectation values of the position operator $\hat{x}$ and its square $\hat{x}^2$.  We choose the Landau gauge, ${\bf A}_0 (\hat{\bf r}) = (0, B \hat{x}, 0)$, for the static vector potential.  This simplifies the calculation because the Hamiltonian becomes translationally invariant in the $y$-direction, and $\hat{p}_y$ is a conserved quantity.

In the Landau gauge, the canonical momenta are $\hat{\pi}_x = \hat{p}_x$ and $\hat{\pi}_y = \hat{p}_y - eB\hat{x}$. We can express $\hat{x}$ in terms of the ladder operators defined in Eq.~\eqref{eq:ladder_ops} and $\hat{p}_y$:
\begin{align}
\hat{x} = \frac{\hat{p}_y}{eB} - i\frac{l_B}{\sqrt{2}}(\hat{a}^\dagger - \hat{a}).
\label{eq:x_ladder_landau_section}
\end{align}
We consider a time-evolved state $\ket{\psi(t)} = \hat{D}(\alpha_t^{(0)})\ket{N'(t)}$, where 
\begin{align}
\ket{N'(t)} = e^{-\frac{i}{\hbar}\int_{t_0}^t \Delta_{t'}^{(0)}dt'}e^{-i \omega_c (\hat{a}^\dagger \hat{a} +1/2) t}\ket{N, k_y} .
\end{align}
Here, $\ket{N, k_y}$ is an eigenstate of the unperturbed Hamiltonian in the Landau gauge, representing the $N$-th Landau level with momentum $\hbar k_y$ in the $y$-direction. Note that because $\hat{p}_y$ is conserved and commutes with the transformed Hamiltonian, its expectation value in the state $\ket{N^{\prime}(t)}$ is simply $\hbar k_y$.

The expectation value of $\hat{x}(t)$ is:
\begin{align}
\langle \hat{x}(t) \rangle_N & = \bra{N'(t)} \hat{D}^\dagger(\alpha_t^{(0)}) \hat{x} \hat{D}(\alpha_t^{(0)}) \ket{N'(t)} \nonumber \\
&= \frac{\hbar k_y}{eB} - \sqrt{2}l_B {\rm Im}[ \alpha_t^{(0)} ].
\label{eq:x_expect_landau_section}
\end{align}
where we used the displacement operator identities in Eq.~\eqref{eq:transformed_ops}.
Substituting the resonant solution for $\alpha_t^{(0)}$, Eq.~\eqref{eq:alpha_resonance_solution}, into Eq.~\eqref{eq:x_expect_landau_section} gives:
\begin{align}
\langle \hat{x}(t) \rangle_N = \frac{\hbar k_y}{eB} - \sqrt{2} l_B \Big( t \frac{z_+}{\hbar}\cos(\omega_c t) +  \frac{z_-}{\hbar \omega_c} \sin(\omega_c t) \Big) .
\end{align}
Similarly, the expectation value of $\hat{x}^2(t)$ is:
\begin{align}
\langle \hat{x}^2(t) \rangle_N = \langle \hat{x}(t) \rangle^2 + l_B^2 (N + 1/2).\label{eq:x_squared_expect_landau_section}
\end{align}
The expectation value $\langle \hat{x}(t) \rangle_N$ contains a term linear in $t$, indicating a drift in the $x$-direction.  Meanwhile, $\langle \hat{x}^2(t) \rangle_N$ contains a term proportional to $t^2$, demonstrating the unbounded spreading of the wavefunction. This confirms the delocalization of the electron and the breakdown of the Floquet picture at resonance when $z_+ \neq 0$.

Despite this delocalization, our exact algebraic approach reveals subtleties in the resonant regime: the distinction between the unbounded growth of the displacement operator and the nature of the energy spectrum.

The unbounded growth of $\alpha_t^{(0)}$ {\it precludes} the existence of conventional time-periodic Floquet states. However, the energy spectrum's character depends sensitively on the driving field's polarization, as shown by examining the scalar potential $\Delta_t^{(0)}$:
\begin{align}
\Delta_t^{(0)} 
& = \frac{{\rm Re}[z_{+}z_{-}^*]+[\cos(2\omega_c t)-1]|z_-|^2}{2 \hbar\omega_c}
\nonumber\\
& -
\frac{{\rm Re}[z_{+} z_{-}^* e^{-2i \omega_c t} (1 + 2 i \omega_c t)]}{2 \hbar\omega_c}
+
\frac{e^2{\bf A}^2 (t)}{2m}.
\label{Delta0_resonance_solution}
\end{align}

For circularly polarized light, where $A_x = A_y = A_0$ and $\phi_y = \pm \pi/2$, we have [see Eqs.~\eqref{eq:driving_term} and \eqref{eq:driving_term_floquet}]:
\begin{align}
\begin{aligned}
\phi_y = +\pi/2: 
\quad
z_+ = 0, 
\quad
z_- = + i \sqrt{\frac{\hbar \omega_c}{2 m}} e A_0 ,
\\
\phi_y = -\pi/2: 
\quad
z_- = 0, 
\quad
z_+ = - i \sqrt{\frac{\hbar \omega_c}{2 m}} e A_0 .
\end{aligned}
\end{align}

This polarization-dependence yields a result that refines Popov's~\cite{popov1970parametric} earlier analysis and provides a more nuanced understanding: when $z_+ = 0$ or $z_- = 0$ (circular polarization), the linear-in-time growth in $\Delta_t^{(0)}$ vanishes, allowing the system to maintain discrete energy levels in the transformed frame despite the unbounded growth in phase space. In contrast, for elliptical or linear polarization ($z_+ z_-^* \neq 0$), the term linear in $t$ appears in $\Delta_t^{(0)}$, producing an explicit continuous energy spectrum.

\subsection{Instantaneous Energy Absorption and Initial State Dependence}
\label{subsec:energy-absorption}

To gain further insight into the driven dynamics, we calculate the instantaneous energy absorption rate, $P(t)$. This quantity is relevant to experiments that probe the energy transfer between the driving field and the 2DEG, and provides information about the time-resolved energy flow.

The instantaneous energy absorption rate is defined as the expectation value of the time derivative of the system Hamiltonian, $\hat{H}(t)$:
\begin{align}
P(t) &= \frac{d}{dt} \bra{\psi(t)} \hat{H}(t) \ket{\psi(t)} = \bra{\psi(t)} \frac{\partial \hat{H}(t)}{\partial t} \ket{\psi(t)},
\label{eq:power_def}
\end{align}
where we applied the Hellmann–Feynman theorem.
Using the explicit form of the Hamiltonian from Eq.~\eqref{eq:hamiltonian_ladder}, we obtain the time derivative:
\begin{align}
\frac{\partial \hat{H}(t)}{\partial t} = - \hat{a} \dot{z}_t^* - \hat{a}^\dagger \dot{z}_t + \frac{e^2}{m} \mathbf{A}(t) \cdot \dot{\mathbf{A}}(t),
\label{eq:dHdt}
\end{align}
where $z_t$ is the complex parameter representing the classical driving field through Eq.~\eqref{eq:driving_term}.

To evaluate $P(t)$, we use the exact solution in the original frame, $\ket{\psi(t)} = \hat{D}(\alpha_t^{(0)})\ket{\psi'(t)}$. The instantaneous power absorption rate becomes:
\begin{align}
P(t) &= \bra{\psi(t)} \frac{\partial \hat{H}(t)}{\partial t} \ket{\psi(t)} \nonumber \\
&= \bra{\psi'(t)} \hat{D}^\dagger(\alpha_t^{(0)}) \frac{\partial \hat{H}(t)}{\partial t} \hat{D}(\alpha_t^{(0)}) \ket{\psi'(t)}.
\label{eq:power_transformed}
\end{align}
Using the displacement operator identities in Eq.~\eqref{eq:transformed_ops}, we transform Eq.~\eqref{eq:dHdt} to the $\hat{H}'(t)$ frame:
\begin{align}
\hat{D}^\dagger(\alpha_t^{(0)}) & \frac{\partial \hat{H}(t)}{\partial t} \hat{D}(\alpha_t^{(0)})
= - \hat{a} \dot{z}_t^* - \hat{a}^\dagger \dot{z}_t 
\nonumber \\
& - \alpha_t^{(0)} \dot{z}_t^* - (\alpha_t^{(0)})^* \dot{z}_t + \frac{e^2}{m} \mathbf{A}(t) \cdot \dot{\mathbf{A}}(t).
\label{eq:dHdt_transformed}
\end{align}

Substituting Eq.~\eqref{eq:dHdt_transformed} into Eq.~\eqref{eq:power_transformed}, we obtain:
\begin{align}
P(t) = &- \dot{z}_t^* \bra{\psi'(t)} \hat{a} \ket{\psi'(t)} - \dot{z}_t \bra{\psi'(t)} \hat{a}^\dagger \ket{\psi'(t)} \nonumber \\
&- \alpha_t^{(0)} \dot{z}_t^* - (\alpha_t^{(0)})^* \dot{z}_t + \frac{e^2}{m} \mathbf{A}(t) \cdot \dot{\mathbf{A}}(t).
\label{eq:power_general}
\end{align}

We now analyze the absorption rate for three different initial states: a Fock state, a coherent state, and a thermal state. For an initial Fock state $\ket{\psi'(0)} = \ket{N}$, we have $\bra{N} \hat{a} \ket{N} = \bra{N} \hat{a}^\dagger \ket{N} = 0$. For an initial coherent state $\ket{\psi'(0)} = \ket{\beta}$, we have $\bra{\beta} \hat{a} \ket{\beta} = \beta$ and $\bra{\beta} \hat{a}^\dagger \ket{\beta} = \beta^*$. And for a thermal state at inverse temperature $1/(k_B T)$, the density matrix is:
\begin{align}
\hat{\rho}_\text{th} = [1 - e^{-\hbar\omega_c/k_B T}] \sum_{N=0}^{\infty} e^{-N\hbar\omega_c/(k_B T)} \ket{N}\bra{N},
\label{eq:thermal_state}
\end{align}
and the thermal averages of the ladder operators vanish:
\begin{align}
\langle\hat{a}\rangle_\text{th} = \text{Tr}(\hat{\rho} \hat{a}) = 0, \quad
\langle\hat{a}^\dagger\rangle_\text{th} = \text{Tr}(\hat{\rho} \hat{a}^\dagger) = 0.
\label{eq:thermal_averages}
\end{align}
Substituting these expectation values into Eq.~\eqref{eq:power_general}, we obtain the following results for each state:
\begin{align}
P_N(t) & = P_\text{th}(t) 
\nonumber \\
& = - \alpha_t^{(0)} \dot{z}_t^* - (\alpha_t^{(0)})^* \dot{z}_t + \frac{e^2}{m} \mathbf{A}(t) \cdot \dot{\mathbf{A}}(t), \label{eq:power_fock_thermal} \\
P_\beta(t) &= - \dot{z}_t^* \beta - \dot{z}_t \beta^* - \alpha_t^{(0)} \dot{z}_t^* - (\alpha_t^{(0)})^* \dot{z}_t \nonumber \\
&\quad + \frac{e^2}{m} \mathbf{A}(t) \cdot \dot{\mathbf{A}}(t). \label{eq:power_coherent}
\end{align}
Notably, the thermal state result $P_\text{th}(t)$ is identical to the Fock state result $P_N(t)$, as both have vanishing expectation values for the ladder operators. The difference between the coherent state and the Fock/thermal states is:
\begin{align}
P_\beta(t) - P_N(t)  = 2 \text{Re}(\beta \dot{z}_t^*).
\label{eq:power_difference}
\end{align}

The instantaneous absorption rate exhibits a  quantum signature in its dependence on initial states. For Fock and thermal states, being eigenstates of the number operator, the absorption derives purely from classical parameters $\alpha_t^{(0)}$ and the external field. Coherent states, in contrast, generate an additional interference term $2 {\rm Re}(\beta \dot{z}_t^*)$ through their phase relationship with the field, which is absent in the number-eigenstate dynamics.

We now focus specifically on the resonant driving case ($\Omega = \omega_c$), where the energy absorption rates are expected to be most pronounced. Inserting Eqs.~\eqref{eq:z_t_resonant} and \eqref{eq:alpha_resonance_solution} into Eq.~\eqref{eq:power_fock_thermal}, we obtain:
\begin{align}
\begin{aligned}
P_N(t) & = P_\text{th}(t) 
\\
& = 
\frac{|z_+|^2}{\hbar} \, 2 \omega_c t - \frac{|z_-|^2}{\hbar} \sin(2 \omega_c t)
\\
& +
\frac{{\rm Re}[z_+ z_-^* (1+2 i \omega_c t)]}{\hbar}
\sin (2 \omega_c t)
\\
& 
- \frac{ {\rm Im}[z_+ z_-^*(1+2i\omega_c t)]}{\hbar} \cos(2 \omega_c t)
\\
& + \frac{{\rm Im}[z_+ z_-^*]}{\hbar} + \frac{e^2}{m} \mathbf{A}(t) \cdot \dot{\mathbf{A}}(t).
\label{eq:power_resonat_final}
\end{aligned}
\end{align}
At resonance, the energy absorption rate exhibits distinctive behavior determined by the driving field polarization. The key feature is a continuous linear growth term proportional to $(|z_+|^2 / \hbar) 2\omega_c t$, reflecting the resonant coupling between the cyclotron motion and the co-rotating component of the drive ($z_+$). This unbounded growth, characteristic of resonant driving in the absence of relaxation, represents continuous energy absorption from the driving field. The remaining terms in Eq.~\eqref{eq:power_resonat_final} oscillating at $2\omega_c$ arise from interference between counter-rotating field components.

The polarization of the driving field controls this behavior through $z_\pm$. For right-circular polarization ($\phi_y = +\pi/2$), $z_+ = 0$ eliminates the linear growth, yielding purely oscillatory absorption. Left-circular polarization ($\phi_y = -\pi/2$) maximizes the growth term through $z_- = 0$. Linear and elliptical polarizations produce both effects through finite $z_\pm$, manifesting as linear growth modulated by $2\omega_c$ oscillations.

\subsection{General Solution for the Displacement Parameter}
\label{subsec:general_alpha}

While specific driving fields, such as those considered in Sections \ref{subsec:non-resonant-Floquet-2DEG} and \ref{subsec:resonant-floquet-2deg}, yield particular closed-form solutions for $\alpha_t^{(0)}$, a general solution to Eq. \eqref{eq:alpha_condition} provides deeper insight into the relationship between an arbitrary driving term $z_t$ and the system's response.

Let us restate Eq. \eqref{eq:alpha_condition}:
\begin{align}
i\hbar\dot{\alpha}_t^{(0)} - \hbar \omega_c \alpha_t^{(0)} = -z_t.
\label{eq:alpha_condition_repeat}
\end{align}
This first-order linear inhomogeneous differential equation can be solved using the method of integrating factors. We multiply both sides by $e^{i\omega_c t}$ to obtain:
\begin{align}
\frac{d}{dt} \left( e^{i\omega_c t} \alpha_t^{(0)} \right) = \frac{i}{\hbar}e^{i\omega_c t} z_t.
\end{align}
Integrating from an initial time $t_0$ to $t$ and assuming the common initial condition where the system starts at rest, i.e., $\alpha_{t_0}^{(0)} = 0$, yields:
\begin{align}
\alpha_t^{(0)} =  \frac{i}{\hbar} \int_{t_0}^t e^{-i\omega_c (t-t')} z_{t'} dt'.
\label{eq:general_alpha_solution_rest}
\end{align}
This solution provides a framework for analyzing various driving protocols beyond the periodic cases discussed earlier. The integral form in Eq.~\eqref{eq:general_alpha_solution_rest} reveals how the displacement parameter accumulates contributions from the driving field at different times. While this integral may not always yield analytical closed forms, it serves as a starting point for more complex driving scenarios such as pulsed or chirped fields.

\subsection{Application to a Constant Electric Field}
\label{subsec:constant_E}

Our formalism can also be readily applied to non-periodic driving scenarios. A fundamental example is a 2DEG subject to a constant, in-plane electric field $\mathbf{E} = (E_x, E_y)$ in the presence of the static perpendicular magnetic field $\mathbf{B}$. This physical situation is central to magnetotransport phenomena and the Hall effect.

A constant electric field can be described by a vector potential that is linear in time, $\mathbf{A}(t) = -\mathbf{E}t$. The complex driving term $z_t$ from Eq.~\eqref{eq:driving_term} consequently takes the form:
\begin{align}
z_t = \left( -e \sqrt{\frac{\hbar \omega_c}{2 m}} (E_x - iE_y) \right) t \equiv Z_E t,
\end{align}
where $Z_E$ is a complex constant determined by the electric field. To find the corresponding displacement parameter $\alpha_t^{(0)}$, we use the general integral solution from Eq.~\eqref{eq:general_alpha_solution_rest}. Assuming the system starts from rest at $t_0=0$ (i.e., $\alpha_{t=0}^{(0)} = 0$), we have:
\begin{align}
\alpha_t^{(0)} =  \frac{i}{\hbar} \int_{0}^t e^{-i\omega_c (t-t')} (Z_E t') dt'.
\end{align}
Evaluating this integral yields the exact displacement parameter:
\begin{align}
\alpha_t^{(0)} = \frac{Z_E}{\hbar\omega_c} t + \frac{i Z_E}{\hbar\omega_c^2} (1 - e^{-i\omega_c t}).
\label{eq:alpha_solution_constE_app}
\end{align}
The physical motion of the electron is revealed by the expectation value of its position. As derived in Eq.~\eqref{eq:x_expect_landau_section}, the position expectation value for an initial Landau state $\ket{N, k_y}$ in the Landau gauge is given by:
\begin{align}
\langle \hat{x}(t) \rangle = \frac{\hbar k_y}{eB} + \sqrt{2} l_B \text{Im}[\alpha_t^{(0)}].
\end{align}
Substituting our solution for $\alpha_t^{(0)}$, the term linear in time (the drift) is $\sqrt{2} l_B \text{Im}[\frac{Z_E}{\hbar\omega_c} t]$. The drift velocity is therefore:
\begin{align}
v_{dx} = \frac{d}{dt}\left( \sqrt{2} l_B \text{Im}\left[\frac{Z_E t}{\hbar\omega_c}\right] \right) = \frac{E_y}{B}.
\end{align}
A similar analysis for $\langle \hat{y}(t) \rangle$ yields $v_{dy} = -E_x/B$. This is precisely the classical Hall drift velocity, $\mathbf{v}_d = (\mathbf{E} \times \mathbf{B})/B^2$.

This drift motion is the time-domain signature of the underlying physics. In a reference frame moving with velocity $\mathbf{v}_d$, the electric field vanishes and the electron dynamics are governed by an effective magnetic field $B' = B\sqrt{1-v_d^2/c^2}$. The ``collapse'' of the Landau level structure occurs at the critical field strength $E = cB$ (where $v_d=c$), at which point $B'$ vanishes and the electron undergoes unbounded acceleration. Our non-relativistic solution captures the onset of this behavior through the linear drift of the position expectation value, which remains a valid description in the limit $E \ll cB$.

\section{Summary and Discussion}

\label{sec:discussion}

Our work introduces an algebraic solution for driven Landau levels in two-dimensional electron gases that remains valid under arbitrary time-dependent driving conditions. The key innovation lies in our use of the displacement operator for time-dependent unitary transformation, which converts the complex time-dependent Hamiltonian into an exactly solvable form consisting of a time-independent harmonic oscillator and a time-dependent scalar potential.

Our formalism, built upon ladder operators, yields clear insights into two important regimes. In the case of non-resonate periodic driving, it provides a natural framework for deriving Floquet states and quasienergies. It also illuminates the resonant driving regime ($\Omega = \omega_c$) where traditional Floquet analysis fails. Furthermore, our analysis of instantaneous energy absorption revealed a quantum interference effect, manifesting as distinct absorption patterns between coherent and Fock/thermal initial states.

The gauge- and representation-independent nature of our solution provides a convenient basis for investigating more realistic systems. This framework extends to scenarios involving electron-electron interactions, disorder effects, and coupling to thermal baths, offering tools to examine phenomena such as zero-resistance states~\cite{mani2002zero,zudov2003evidence} and to uncover the underlying negative resistivity~\cite{andreev2003dynamical} that gives rise to these states.

\begin{acknowledgments}
The author gratefully acknowledges Inti Sodemann for enlightening discussions, Jie Wang for the helpful conversations at the Max Planck Institute for the Physics of Complex Systems (MPIPKS), and Oles Matsyshyn for the stimulating exchanges.
\end{acknowledgments}

\bibliography{reference}

\appendix

\section{Derivative of the Displacement Operator}
\label{app:displacement_derivative}

In this appendix, we provide a detailed derivation of the time derivative of the displacement operator. We begin by recalling the definition of the displacement operator:
\begin{align}
\hat{D}(\alpha_t) &= \exp(\alpha_t \hat{a}^\dagger - \alpha_t^* \hat{a}), 
\label{eq:app_displacement_def}
\end{align}
where $\alpha_t$ is a time-dependent complex function.
It transforms the ladder operators according to:
\begin{align}
\begin{aligned}
& \hat{D}^\dagger(\alpha_t)\hat{a}\hat{D}(\alpha_t) = \hat{a} + \alpha_t, \\
& \hat{D}^\dagger(\alpha_t)\hat{a}^\dagger \hat{D}(\alpha_t) = \hat{a}^\dagger + \alpha_t^*.
\end{aligned}
\end{align}
Therefore
\begin{align}
\hat{D}^\dagger(\alpha_t) \hat{a}^\dagger \hat{a} \hat{D}(\alpha_t)
=
( \hat{a}^\dagger + \alpha_t^* )
( \hat{a} + \alpha_t ).
\end{align}
To compute the time derivative, we first define the exponent operator:
\begin{align}
\hat{\theta}(t) &= \alpha_t\hat{a}^\dagger - \alpha_t^*\hat{a},
\label{eq:app_theta_def}
\end{align}
whose time derivative is:
\begin{align}
\dot{\hat{\theta}}(t) &= \dot{\alpha}_t\hat{a}^\dagger - \dot{\alpha}_t^*\hat{a}.
\label{eq:app_theta_dot}
\end{align}
A key step is to evaluate the commutator between $\hat{\theta}(t)$ and $\dot{\hat{\theta}}(t)$:
\begin{align}
[\hat{\theta}(t), \dot{\hat{\theta}}(t)] &= [\alpha_t\hat{a}^\dagger - \alpha_t^*\hat{a}, \dot{\alpha}_t\hat{a}^\dagger - \dot{\alpha}_t^*\hat{a}] \nonumber \\
&= -(\alpha_t\dot{\alpha}_t^* - \alpha_t^*\dot{\alpha}_t) \nonumber \\
&= -2i\,\text{Im}(\alpha_t\dot{\alpha}_t^*).
\label{eq:app_commutator}
\end{align}
To evaluate the time derivative of the exponential operator, we use:
\begin{align}
\frac{\partial}{\partial t}\hat{D}(\alpha_t) &= \lim_{\epsilon \to 0} \frac{\hat{D}(\alpha_{t+\epsilon}) - \hat{D}(\alpha_t)}{\epsilon} \nonumber \\
&= \lim_{\epsilon \to 0} \frac{e^{\hat{\theta}(t) + \epsilon\dot{\hat{\theta}}(t) + O(\epsilon^2)} - e^{\hat{\theta}(t)}}{\epsilon}.
\label{eq:app_derivative_limit}
\end{align}
Since $[\hat{\theta}(t), \dot{\hat{\theta}}(t)]$ is a scalar, we can apply the Baker-Campbell-Hausdorff formula:
\begin{align}
e^{\hat{A}+\hat{B}} = e^{\hat{A}}e^{\hat{B}}e^{-\frac{1}{2}[\hat{A},\hat{B}]},
\label{eq:app_bch}
\end{align}
valid when $[\hat{A},[\hat{A},\hat{B}]] = [\hat{B},[\hat{A},\hat{B}]] = 0$. Setting $\hat{A} = \hat{\theta}(t)$ and $\hat{B} = \epsilon\dot{\hat{\theta}}(t)$, and expanding the exponential to first order in $\epsilon$, we obtain:
\begin{align}
\frac{\partial}{\partial t} \hat{D}(\alpha_t) &= \hat{D}(\alpha_t)\Big(\dot{\hat{\theta}}(t) - \frac{1}{2}[\hat{\theta}(t), \dot{\hat{\theta}}(t)]\Big).
\label{eq:app_derivative_final}
\end{align}
Finally, substituting the explicit forms of $\dot{\hat{\theta}}(t)$ and the commutator:
\begin{align}
-i\hbar\hat{D}^\dagger(\alpha_t)\frac{\partial}{\partial t} \hat{D}(\alpha_t) &= -i\hbar\dot{\alpha}_t\hat{a}^\dagger + i\hbar\dot{\alpha}_t^*\hat{a} \nonumber \\
&\quad - \hbar\,\text{Im}(\alpha_t\dot{\alpha}_t^*),
\label{eq:app_transformed_derivative}
\end{align}
which is the form used in the main text to derive the transformed Hamiltonian.

\end{document}